\documentclass[apjl]{emulateapj}
\usepackage{graphicx}
\usepackage{amsmath}
\usepackage{natbib}
\usepackage[table]{xcolor}
\usepackage{multirow}
\usepackage{rotating}
\usepackage{tabu}

\usepackage[utf8x]{inputenc}
\usepackage[table]{xcolor}
\usepackage{tabu}
\usepackage{ulem}

\newcommand\simlt{\lower.5ex\hbox{$\; \buildrel < \over \sim \;$}}
\newcommand\simgt{\lower.5ex\hbox{$\; \buildrel > \over \sim \;$}}

\colorlet{MyGreen}{green!80!yellow!50}
\colorlet{MyLighterGreen}{green!70!yellow!40}

\colorlet{FreshGray}{gray!20!white}
\colorlet{FreshGreen}{green!20!white}
\colorlet{FreshYellow}{yellow!20!white}
\colorlet{FreshRed}{red!20!white}

\colorlet{white}{gray!0!white}
\colorlet{bands}{gray!20!white}
\colorlet{ModelB}{green!20!white}
\colorlet{FreshYellow}{yellow!20!white}
\colorlet{ModelA}{red!20!white}

\colorlet{modelA}{red!16!white}
\colorlet{lightA}{red!8!white}

\colorlet{modelB}{blue!10!white}
\colorlet{lightB}{blue!70!white!5}

\colorlet{flux}{gray!10!white}
\colorlet{lightFlux}{gray!2!white}

\definecolor{index}{cmyk}{0.1,0.,0.1,0.}
\definecolor{lightIndex}{cmyk}{0.05,0.,0.05,0.}

\begin{document}

\title{Probing the Extragalactic Cosmic Rays origin with gamma-ray and neutrino backgrounds}

\author{Noemie Globus\altaffilmark{1}, Denis Allard\altaffilmark{2}, Etienne Parizot\altaffilmark{2},  Tsvi Piran\altaffilmark{1}}
\altaffiltext{1}{Racah Institute of Physics, The Hebrew University, 91904 Jerusalem, Israel}
\altaffiltext{2}{Laboratoire Astroparticule et Cosmologie, Universit\'e Paris Diderot/CNRS, 10 rue A. Domon et L. Duquet, F-75205 Paris Cedex 13, France}

\begin{abstract}
GeV-TeV gamma-rays and PeV-EeV neutrino backgrounds provide a unique window on the nature of the ultra-high-energy cosmic-rays (UHECRs).
We discuss the implications of the recent Fermi-LAT data regarding the extragalactic gamma-ray background (EGB) and related estimates of the contribution of point sources as well as IceCube neutrino data on the origin of the UHECRs. We calculate the diffuse flux of cosmogenic $\gamma$-rays and neutrinos produced by the UHECRs and derive constraints  on the possible cosmological evolution of UHECR sources.  In particular, we show that the mixed-composition scenario considered in \citet{Globus2015b}, which is in agreement with both (i) Auger measurements of the energy spectrum and composition up to the highest energies and (ii) the ankle-like feature in the light component detected by KASCADE-Grande, is compatible with both the Fermi-LAT measurements and with current IceCube limits. We  also discuss the possibility for future experiments to detect associated cosmogenic neutrinos and further constrain the UHECR models, including possible subdominant UHECR proton sources.

\end{abstract}

\section{Motivation}
\label{sec:motivation}

The interaction of UHECRs with the photon backgrounds during their propagation in intergalactic space produces 
cosmogenic $\gamma$-ray photons \citep{Strong73,Stecker73} through electromagnetic cascades that contribute to the extragalactic gamma-ray background (EGB) at GeV-TeV energies, and cosmogenic neutrinos \citep[$\nu$s, ][]{Bere69} mostly from PeV to multi-EeV energies. 
The flux of these secondary messengers is highly sensitive to the spectral shape, maximal energy, composition and cosmological evolution of the UHECR sources. Therefore one can derive  important constraints on the UHECR origin from a multi-messenger approach that takes these into account
\citep[for $\gamma$-rays]{Protheroe96,CoppAha97,
Ahlers11,2011A&A...535A..66D,Bere16, Supanitsky16,2016ApJ...822...56G}; \citep[e.g.][for $\nu$s]{Stecker79, 
Engel01, Seckel05, Allard06, 
Anchordoqui07, Ahlers09, Kotera10}. 

Source models implying a cosmological evolution much stronger than the star formation rate (SFR) 
have already been 
ruled out as the main UHECR contributors by  
the first Fermi-LAT estimates of the purely diffuse component of the EGB \citep{Fermi10},  
independently of the maximum energy of UHECRs ($E_{\rm max}$), 
in particular for steep (soft) cosmic-ray injection spectra \citep[e.g.][]{Bere10,Ahlers10,2011A&A...535A..66D}. These strong evolutions have also been ruled out by the IceCube limits 
on $\nu$s, in the case of source spectra with large values of the maximum energy-per-nucleon \citep[$E_{\rm max}/A \gtrsim 10^{20}$~eV, see][]{Aartsen16}.

Moreover, the recent Fermi-LAT 
data \citep{2015ApJ...799...86A}, together with 
statistics of the photon counts in the skymap pixels
\citep[e.g.][and references therein]{Malyshev11} have enabled different authors \citep[][hereafter A16 and Z16]{Ack2016,Zechlin16} to estimate the flux contributed by point sources (PS) well below the Fermi-LAT detection limits. 
These studies show that resolved and unresolved PS account for the majority of the EGB. 
Since a $\gamma$-ray background due to  extragalactic cosmic rays (EGCRs) is unavoidable, it is crucial to verify that the proposed UHECR source models do not violate the existing constraints.

Recent measurements by the Pierre Auger Observatory (Auger) 
indicate that the composition of UHECRs is mixed (predominantly light) at the ankle of the cosmic-ray spectrum, and it gets progressively heavier as the energy increases 
 \citep{Aab14b
 }. This composition trend can be interpreted as the signature of a low maximal energy-per-unit-charge ($E_{\rm max}/Z\lesssim  10^{19}$~eV) 
of the nuclei accelerated at the dominant sources of UHECRs. 
Below $10^{18}$ eV, the KASCADE-Grande experiment 
reported an ankle-like feature in the energy spectrum of light (proton-helium) elements with a break at $\sim 10^{17}$~eV \citep{Apel13,Bertaina15}.
This 
``light ankle'' can be naturally understood as the emergence of a light EGCR component, 
taking over the steeper Galactic cosmic-ray (GCR) component. 

In this Letter, we investigate  constraints that can be set on mixed-composition EGCR models, taking into account the most recent Fermi-LAT estimates  of the EGB and its unresolved component. 
We discuss the viability of a class of mixed-composition models in which the KASCADE-Grande and Auger data are understood in terms of a transition between a GCR component and a single EGCR component with a soft proton spectrum and low $E_{\rm max}$. 
This soft proton component would be responsible for the light ankle and it would be the dominant contributor to the cosmogenic $\gamma$-ray flux.
This model was shown to be compatible with the spectrum and composition data at all energies \citep[][hereafter G15b]{Globus2015b}, and it is consistent with the anisotropy constraints on galactic protons \citep{Tinyakov16}.


\section{\label{sec:model}Source model}

Any phenomenological EGCR model that account for the data needs a very hard spectrum at the sources, to reproduce the evolution of the composition above the ankle observed by Auger, and a soft proton component, to account for the light ankle seen by KASCADE-Grande.
As an example we consider the  EGCR source model for UHECR acceleration at gamma-ray bursts (GRB) internal shocks \citep[][hereafter G15a]{Globus2015a}, whose basic features result from the presence of a dense, broad-band photon field in the acceleration environment, and should thus also be expected in other types of powerful high-energy sources.
Those features are:\\
$-$ A very hard source spectrum for the composed nuclei (harder than $\sim~E^{-1}$ below  $E_{\rm max}(Z)$), with a rigidity-dependent cut-off due to the selection of high rigidity particles by the escape process. \\
$-$ A much softer source spectrum for the nucleons, due to the free escape of neutrons produced by the photo-disintegration of nuclei.\\ 
Both features would arise in any model based on electromagnetic acceleration including a significant dissociation of the nuclei at the source.

\begin{figure}[ht!]
 \centering
  \includegraphics[width=\linewidth]{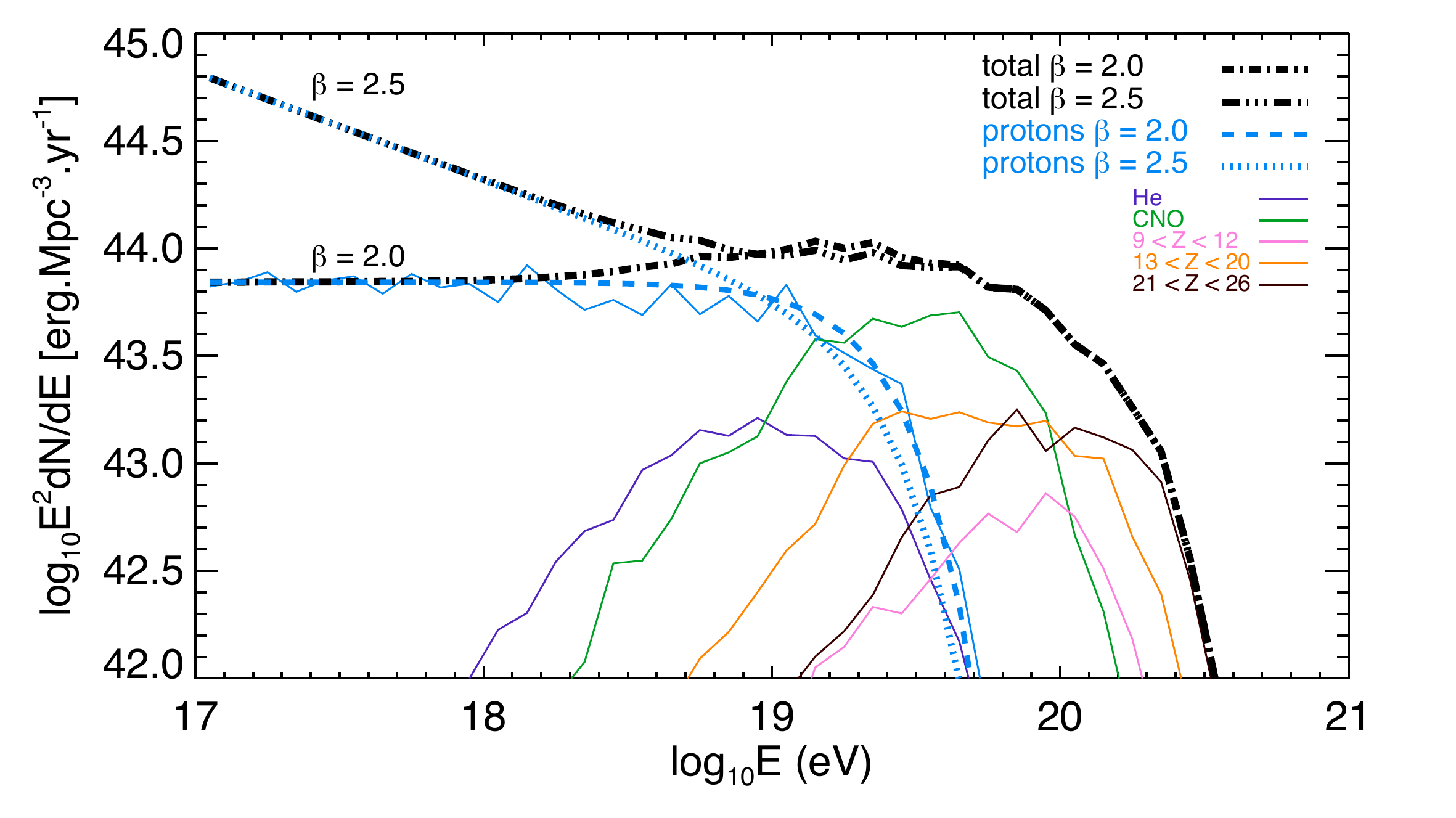}
  \includegraphics[width=\linewidth]{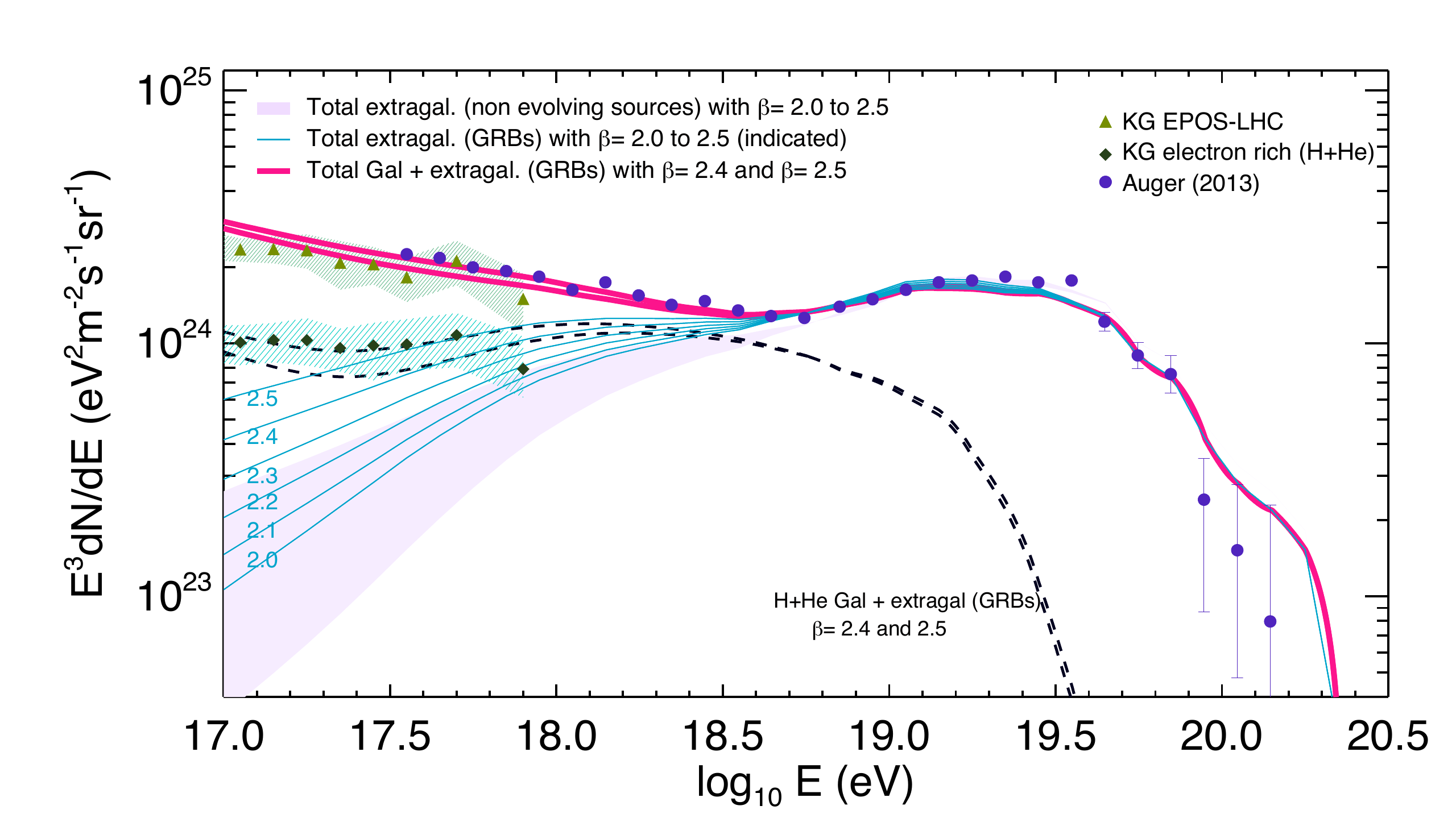}
 \caption{Upper panel: UHECR injection spectrum for the various nuclei, as obtained in G15a, with the fit of the proton component with spectral index $\beta = 2.0$ (dashed blue), and with its modified shape in the case of $\beta=2.5$. Lower panel: Propagated UHECR spectra  for $2.0\leq\beta\leq2.5$, compared to KASCADE-Grande and Auger data, for a GRB-like cosmological evolution (blue lines) or non-evolving sources (violet shaded area). The total (GCR+EGCR) light component  is compared to that deduced from KASCADE-Grande data (using the EPOS-LHC \citep{EPOS1} hadronic model), for GRB-like evolution with $\beta=2.4$ and 2.5 (dashed lines).}
\label{fig:modelc} 
\end{figure}

The exact shape of the source spectrum of the escaping nucleons 
and composed nuclei depends on various physical parameters, such as  
 the shock 
 geometry and its time evolution, the local magnetic turbulence, 
and the competition between 
energy losses and escape 
 (G15a). 
 Moreover, the distribution of source luminosities influences the shape of the {\it effective}  UHECR 
 spectrum (obtained after convoluting the individual source spectra by the source luminosity function).
The effective spectrum from the GRB model (G15a) is displayed in the upper panel of Fig.~\ref{fig:modelc}.


Since 
the extragalactic protons around $10^{17}$~eV contribute significantly to 
the expected cosmogenic $\gamma$-ray flux in the Fermi energy range, 
we explore, for the sake of generality, (i) different slopes for the  proton component  
(as could result from different physical parameters describing the sources) 
while keeping the same maximal rigidity  and 
spectral shape for heavier nuclei; 
 (ii) different cosmological evolutions, assuming an average source power proportional to $(1+z)^{\alpha}$ up to a redshift $z_{\max}$.

 The soft proton component of the effective UHECR spectrum (upper panel of Fig.~\ref{fig:modelc})  is well fitted by a power law with a Gaussian cut-off, $\mathrm{d}N/\mathrm{d}E \propto E^{-\beta}\exp[-E^2/(2E_{\max}^2)]$ with $\beta=2.0$ and $E_{\max}~\simeq~1.7 \times 10^{19}$~eV. 
 In the following, we allow for a modification of the original proton spectrum, and  consider a range of spectral indices  $2.0\leq\beta\leq2.5$.  The two proton spectra  obtained with the  extreme values of $\beta$  are represented  by thick dashed and dotted blue lines, respectively. The implied range of UHECR emissivities above $10^{17}$~eV is  $L_{\rm CR}^{17}\sim[5.7-14]\cdot10^{44}\,\mathrm{erg}\,\,\mathrm{Mpc}^{-3}\,\, \mathrm{yr}^{-1}$. 
When considering different cosmological evolutions, we need to further rescale the propagated spectrum by a factor between $\sim 0.8$ and $\sim 1.5$ to match the Auger data at high energy.
The Monte-Carlo procedure used to calculate the cosmic-ray, $\nu$ and $\gamma$-ray spectra is presented in \citet{2011A&A...535A..66D}. 

\section{\label{uhecr}Propagated cosmic-ray spectra} 

The lower panel of Fig.~\ref{fig:modelc}, depicts the propagated UHECR spectra for $2.0\leq\beta\leq2.5$,
for EGCR sources evolving as GRBs \citep[][blue lines]{2010MNRAS.406.1944W} and for non evolving sources (violet shaded area). 
Varying the cosmological evolution of UHECR sources does not affect the high-energy part of the propagated spectrum, since the sources contributing at these energies are located at low redshifts (due to the GZK horizon effect). However, a stronger source evolution implies a larger contribution of the more distant sources and thus a larger UHECR flux at lower energies. As a result, a suitable combination of the soft proton source spectrum and a strong cosmological evolution can reproduce the light (supposedly proton-helium) cosmic-ray component estimated from KASCADE-Grande data.

In the case of a GRB-like cosmological evolution (or SFR-like \citep{Yuksel08} that gives very similar results), proton spectral indices $\beta \simeq 2.4-2.5$ provide a good fit to the KASCADE-Grande data when summing the light EGCR component with the GCR light component obtained in G15b (dashed line in Fig.~\ref{fig:modelc}). The resulting proton abundance increases over the $10^{17}-10^{18}$~eV energy range, before slowly dropping above the ankle, reproducing the observed composition trend in the GCR-to-EGCR transition and above. 

In a non-evolving scenario, softer proton indices ($\beta\sim2.7$, and thus larger injection power density) are required to obtain such a large contribution of the EGCR component at low energy. 
Conversely, a stronger source evolution than that of GRBs would require harder proton indices.

\renewcommand{\arraystretch}{1.65}
\begin{table*}[ht!]
\centering
\hspace{50pt}
\tabulinesep=1.2mm

\begin{tabular}{*{7}{|c}|}
\hline
\rowcolor{flux}
\vspace{-2.3pt}
\textbf{Energy bands}
& \textcircled{\raisebox{-.9pt}{1}}
&  \textcircled{\raisebox{-.9pt}{2}}
&  \textcircled{\raisebox{-.9pt}{3}}
& \textcircled{\raisebox{-.9pt}{4}}
&  \textcircled{\raisebox{-.9pt}{5}}
&  \textcircled{\raisebox{-.9pt}{6}}\\

\rowcolor{flux}
(in GeV) & 1.04--1.99 & 1.99--5.0 & 5.0--10.4 & 10.4--50 & 50--171 & 50--2000 \\
\hline
$F_{\rm PS}$ ($\times10^{-9}\,\mathrm{cm}^{-2}\cdot\mathrm{s^{-1}}\cdot\mathrm{sr^{-1}}$) & $250^{+20}_{-40}$ & $124^{+7}_{-25}$ & $27^{+8}_{-3}$ & $14^{+6}_{-1}$ & $1.7^{+1.1}_{-0.4}$ & $2.07^{+0.40}_{-0.34}$ \\
\hline
$F_{\rm PS}/F_{\rm EGB}$ (\% Model A) & $83^{+7}_{-13}$ & $79^{+4}_{-16}$ & $66^{+20}_{-7}$ & $66^{+28}_{-5}$ & $81^{+52}_{-19}$ & $86^{+16}_{-14}$ \\
\hline
$F_{\rm PS}/F_{\rm EGB}$ (\% Model B) & $68^{+5}_{-10}$ & $63^{+4}_{-13}$ & $52^{+15}_{-6}$ & $51^{+22}_{-4}$ & $65^{+41}_{-15}$ & $71^{+13}_{-12}$ \\
\hline
\hline
$F_{\rm SFG+misAGN}$ ($\times10^{-9}\,\mathrm{cm}^{-2}\cdot\mathrm{s^{-1}}\cdot\mathrm{sr^{-1}}$) & $94^{+100}_{-36}$ & $44^{+49}_{-18}$ & $10^{+12}_{-4}$ & $4.5^{+5.4}_{-1.9}$ & $0.17^{+0.18}_{-0.07}$ & $0.18^{+0.19}_{-0.07}$ \\
\hline
$F_{\rm SFG+misAGN}/F_{\rm EGB}$ (\% Model A) & $31^{+33}_{-12}$ & $28^{+31}_{-11}$ & $25^{+30}_{-10}$ & $21^{+25}_{-9}$ & $8^{+9}_{-3}$ & $7^{+8}_{-3}$ \\
\hline
$F_{\rm SFG+misAGN}/F_{\rm EGB}$ (\% Model B) & $25^{+27}_{-10}$ & $23^{+25}_{-9}$ & $20^{+23}_{-8}$ & $16^{+20}_{-7}$ & $6^{+7}_{-3}$ & $6^{+6}_{-2}$ \\
\hline
\end{tabular}

\caption{Integrated $\gamma$-ray fluxes of PS, estimated by \citet{Zechlin16} in energy bands 1 to 5, and by \citet{Ack2016} in energy band 6, and of the SFG+misAGN components (see text), as modelled by \citet{Ack2012} and \citet{Inoue11}, respectively. The corresponding relative contributions to the total EGB flux is also given in percent, assuming Galactic foreground models A or B \citep{2015ApJ...799...86A}.}
\label{table1}
 \renewcommand{\arraystretch}{1.4}


\hspace{20pt}
\centering


\begin{tabular}{*{17}{|c}|}
\hline
\multicolumn{3}{|c|}{{{} Components }} & \multicolumn{6}{|c|}{ {{} Energy bands ($\beta = 2.0$)}} & \multicolumn{6}{|c|}{ {{} Energy bands ($\beta = 2.5$)}} \\
\cline{4-15}
\multicolumn{3}{|c|}{ and source evolution
}
& \multicolumn{1}{|c|}{ \textcircled{\raisebox{-.9pt}{1}}}
& \textcircled{\raisebox{-.9pt}{2}}
&  \textcircled{\raisebox{-.9pt}{3}}
& \textcircled{\raisebox{-.9pt}{4}}
&  \textcircled{\raisebox{-.9pt}{5}}
& \textcircled{\raisebox{-.9pt}{6}}
& \multicolumn{1}{|c|}{ \textcircled{\raisebox{-.9pt}{1}}}
&  \textcircled{\raisebox{-.9pt}{2}}
& \textcircled{\raisebox{-.9pt}{3}}
& \textcircled{\raisebox{-.9pt}{4}}
&  \textcircled{\raisebox{-.9pt}{5}}
& \textcircled{\raisebox{-.9pt}{6}}
\\
\hline
\hline

\rowcolor{flux}
\multicolumn{2}{|c|}{{} $F_{\rm UHECR}$} & GRB & \cellcolor{lightFlux} 170 &\cellcolor{lightFlux} 120 & \cellcolor{lightFlux} 44 & \cellcolor{lightFlux} 32 & \cellcolor{lightFlux} 2.5 & \cellcolor{lightFlux} 2.7 & 260  & 190 & 67 & 48 & 3.4 & 3.7 \\
\rowcolor{flux}
\multicolumn{2}{|c|}{{} {($\times10^{-10}\,\mathrm{cm}^{-2}\,\mathrm{s}^{-1}\,\mathrm{sr}^{-1}$)}} & SFR &\cellcolor{lightFlux} 200 &  \cellcolor{lightFlux}140 & \cellcolor{lightFlux}51 & \cellcolor{lightFlux} 38 &  \cellcolor{lightFlux}3.6 &\cellcolor{lightFlux}  3.9 & 270 & 190 & 70 & 52 & 4.7 & 5.1 \\
\rowcolor{flux}
\multicolumn{2}{|c|}{{} } & non evol & \cellcolor{lightFlux}42 & \cellcolor{lightFlux} 30 & \cellcolor{lightFlux} 11 & \cellcolor{lightFlux}8.6 & \cellcolor{lightFlux} 1.1 &\cellcolor{lightFlux}1.3 & 58 & 41 & 15 & 11 & 1.4 & 1.6 \\
\hline
\hline

\rowcolor{modelB}
& & GRB & \cellcolor{lightB} 4.6 & \cellcolor{lightB} 6.2 & \cellcolor{lightB} 8.5 & \cellcolor{lightB} 12 & \cellcolor{lightB} 9.6 & \cellcolor{lightB} 9.4 & 7.0 & 9.5 & 13 & 17 & 13 & 13 \\
\rowcolor{modelB}
& & SFR & \cellcolor{lightB} 5.3 & \cellcolor{lightB} 7.1 & \cellcolor{lightB} 9.8 & \cellcolor{lightB} 14 & \cellcolor{lightB} 14 & \cellcolor{lightB} 13 & 7.3 & 9.9 & 14 & 19 & 18 & 17 \\
\rowcolor{modelB}
& \multirow{-3}{*}{$F_{\rm UHECR}/F_{\rm EGB}$}
& non evol & \cellcolor{lightB} 1.1 & \cellcolor{lightB} 1.6 & \cellcolor{lightB} 2.2 & \cellcolor{lightB} 3.1 & \cellcolor{lightB} 4.2 & \cellcolor{lightB} 4.4 & 1.6 & 2.1 & 2.9 & 4.1 & 5.3 & 5.4 \\
\cline{2-15}
\cline{2-15}
\cline{2-15}


\rowcolor{modelB}
& & GRB & \cellcolor{lightB} \textbf{97} & \cellcolor{lightB} \textbf{92} & \cellcolor{lightB} \textbf{81} & \cellcolor{lightB} \textbf{79} & \cellcolor{lightB} \textbf{80} & \cellcolor{lightB} \textbf{86} & \textbf{100} & \textbf{95} & \textbf{85} & \textbf{85} & \textbf{84} & \textbf{89} \\
\rowcolor{modelB}
& & SFR & \cellcolor{lightB} \textbf{98} & \cellcolor{lightB} \textbf{93} & \cellcolor{lightB} \textbf{82} & \cellcolor{lightB} \textbf{81} & \cellcolor{lightB} \textbf{85} & \cellcolor{lightB} \textbf{90} & \textbf{100} & \textbf{96} & \textbf{86} & \textbf{86} & \textbf{89} & \textbf{94} \\
\rowcolor{modelB}
\multirow{-6}{*}{\rotatebox{90}{\% Model B}} & \multirow{-3}{*}{\textbf{$\displaystyle\frac{F_{\rm (UHECR+PS+SFG+misAGN)}}{F_{\rm EGB}}$}}
& non evol & \cellcolor{lightB} \textbf{94} & \cellcolor{lightB} \textbf{87} & \cellcolor{lightB} \textbf{74} & \cellcolor{lightB} \textbf{70} & \cellcolor{lightB} \textbf{75} & \cellcolor{lightB} \textbf{81} & \textbf{94} & \textbf{88} & \textbf{75} & \textbf{71} & \textbf{76} & \textbf{82} \\
\hline
\hline

\rowcolor{modelA}
& & GRB & \cellcolor{lightA} 5.7 & \cellcolor{lightA} 7.7 & \cellcolor{lightA} 11 & \cellcolor{lightA} 15 & \cellcolor{lightA} 12 & \cellcolor{lightA} 11 & 8.7 & 12 & 16 & 22 & 16 & 15 \\
\rowcolor{modelA}
& & SFR & \cellcolor{lightA} 6.5 & \cellcolor{lightA} 8.9 & \cellcolor{lightA} 12 & \cellcolor{lightA} 18 & \cellcolor{lightA} 17 & \cellcolor{lightA} 16 & 9.0 & 12 & 17 & 24 & 23 & 21 \\
\rowcolor{modelA}
& \multirow{-3}{*}{$F_{\rm UHECR}/F_{\rm EGB}$}
& non evol &  \cellcolor{lightA} 1.4 &  \cellcolor{lightA} 1.9 &  \cellcolor{lightA} 2.8 &  \cellcolor{lightA} 4.0 &  \cellcolor{lightA} 5.3 &  \cellcolor{lightA} 5.3 & 1.9 & 2.6 & 3.7 & 5.4 & 6.7 & 6.6 \\
\cline{2-15}
\cline{2-15}
\cline{2-15}

\rowcolor{modelA}
& & GRB &  \cellcolor{lightA} \textbf{120} &  \cellcolor{lightA} \textbf{115} &  \cellcolor{lightA} \textbf{102} &  \cellcolor{lightA} \textbf{102} &  \cellcolor{lightA} \textbf{101} &  \cellcolor{lightA} \textbf{105} & \textbf{123} & \textbf{119} & \textbf{108} & \textbf{110} & \textbf{105} & \textbf{108} \\
\rowcolor{modelA}
& & SFR &  \cellcolor{lightA} \textbf{121} &  \cellcolor{lightA} \textbf{116} &  \cellcolor{lightA} \textbf{104} &  \cellcolor{lightA} \textbf{105} &  \cellcolor{lightA} \textbf{107} &  \cellcolor{lightA} \textbf{110} & \textbf{123} & \textbf{120} & \textbf{108} & \textbf{112} & \textbf{112} & \textbf{114} \\
\rowcolor{modelA}
\multirow{-6}{*}{\rotatebox{90}{\% Model A}} & \multirow{-3}{*}{\textbf{$\displaystyle\frac{F_{\rm (UHECR+PS+SFG+misAGN)}}{F_{\rm EGB}}$}}
& non evol &  \cellcolor{lightA} \textbf{116} &  \cellcolor{lightA} \textbf{109} &  \cellcolor{lightA} \textbf{94} &  \cellcolor{lightA} \textbf{91} &  \cellcolor{lightA} \textbf{94} &  \cellcolor{lightA} \textbf{99} & \textbf{116} & \textbf{110} & \textbf{95} & \textbf{93} & \textbf{96} & \textbf{100} \\
\hline
\hline

\rowcolor{modelA}
& & GRB &  \cellcolor{lightA} 89 &  \cellcolor{lightA} 87 &  \cellcolor{lightA} 77 &  \cellcolor{lightA} 81 &  \cellcolor{lightA} 93 &  \cellcolor{lightA} 97 & 92 & 91 & 82 & 88 & 97 & 101 \\
\rowcolor{modelA}
& & SFR &  \cellcolor{lightA} 89 &  \cellcolor{lightA} 88 &  \cellcolor{lightA} 78 &  \cellcolor{lightA} 84 &  \cellcolor{lightA} 98 &  \cellcolor{lightA} 102 & 92 & 91 & 83 & 90 & 104 & 107 \\
\rowcolor{modelA}
\multirow{-3}{*}{\rotatebox{90}{\footnotesize \% Mod  A}} & \multirow{-3}{*}{$F_{\rm (UHECR+PS)}/F_{\rm EGB}$}
& non evol &  \cellcolor{lightA} 84 & \cellcolor{lightA} 81 &  \cellcolor{lightA} 69 &  \cellcolor{lightA} 70 &  \cellcolor{lightA} 86 & \cellcolor{lightA} 91 &   85 &  82 &  70 &  71 &  88 &  93 \\
\hline

\end{tabular}

\caption{EGCR-induced $\gamma$-ray fluxes in the six energy bands of Table~\ref{table1}, as computed with our mixed-composition model and the two extreme spectral indices of the soft proton component, $\beta=2.0$ and $\beta=2.5$, for three different assumptions regarding the cosmological evolution of the sources (GRB, SFR, and non evolving). The corresponding percentage of the total EGB is given, for models A and B, as well as the percentage contributed by the sum of UHECR+misAGN+SFG+PS components (using central values). In the case of model A, the total UHECR+PS is also shown separately.}
\label{table2}
\end{table*}


 \begin{figure} [t]
 \centering
  \includegraphics[width=\linewidth]{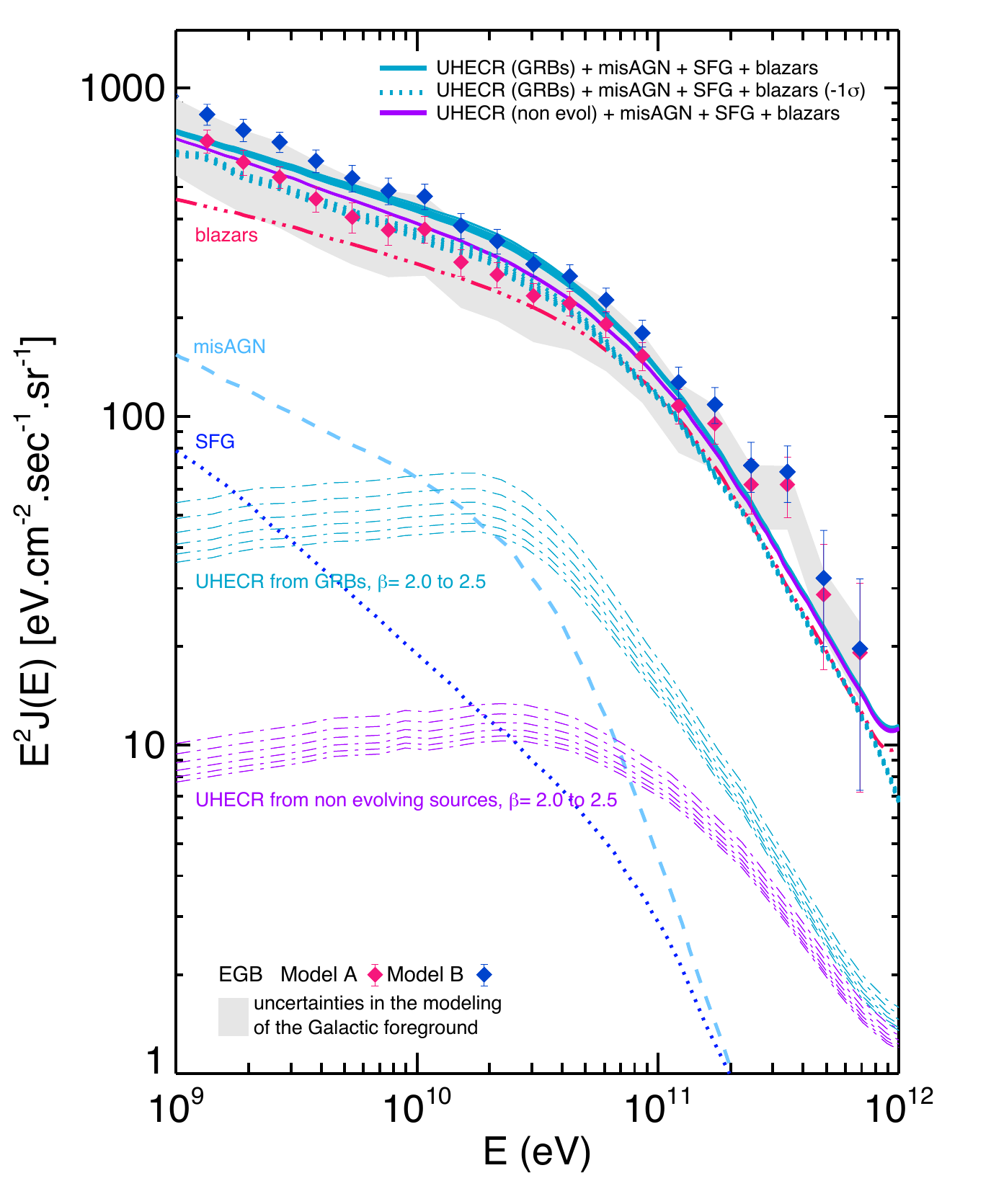}
 \caption{
$\gamma$-ray fluxes from EGCRs (dashed-dotted lines), for GRB-like evolution (blue) and non-evolving (violet) sources, as computed with our mixed-composition model and spectral indices of the soft proton component $2.0\leq\beta\leq2.5$.
Also represented the $\gamma$-ray fluxes from SFG, MisAGN and blazar sources (see labels) as modelled by \cite{Ack2012,Inoue11,Ajello15} respectively. The corresponding sum of UHECR, SFG, misAGN and blazar components is represented by thick solid lines  \citep[or with a dotted line when 1-$\sigma$ lower bound are adopted for the SFG+misAGN+blazar model, see][]{Ajello15}, and compared to the EGB estimated from Fermi-LAT data, for both foreground models A and B.  
 }
 \label{fig:EGB}
 \end{figure}

\section{The  Gamma-ray background}

\label{sec:gamma}
The interactions of the propagating  EGCRs leads to the production of cosmogenic $\gamma$-rays in the GeV-TeV range, through the development of electromagnetic cascades.
The resulting spectra 
are shown  in Fig.~\ref{fig:EGB} for a mixed-composition model with proton spectral indices $2.0\leq\beta\leq2.5$, for sources 
 with no cosmological evolution (violet lines) and with a GRB-like evolution (in blue). 
For a given source evolution, softer proton injection spectra result in larger $\gamma$-ray fluxes, due to the larger amount of low energy protons which efficiently fuel the electromagnetic cascades via the pair production process.
These  $\gamma$-ray fluxes represent only a small contribution to the total EGB, which is reproduced from \citet{2015ApJ...799...86A} for two different models of the Galactic $\gamma$-ray foreground, referred to as model A and model B by the authors, according to whom neither is preferred over the other. 
These two models roughly differ by $\sim 20-30\%$, which can be seen as a rough estimate of their systematics in the subtraction process.

\begin{figure}[t]
\centering
\includegraphics[width=\linewidth]{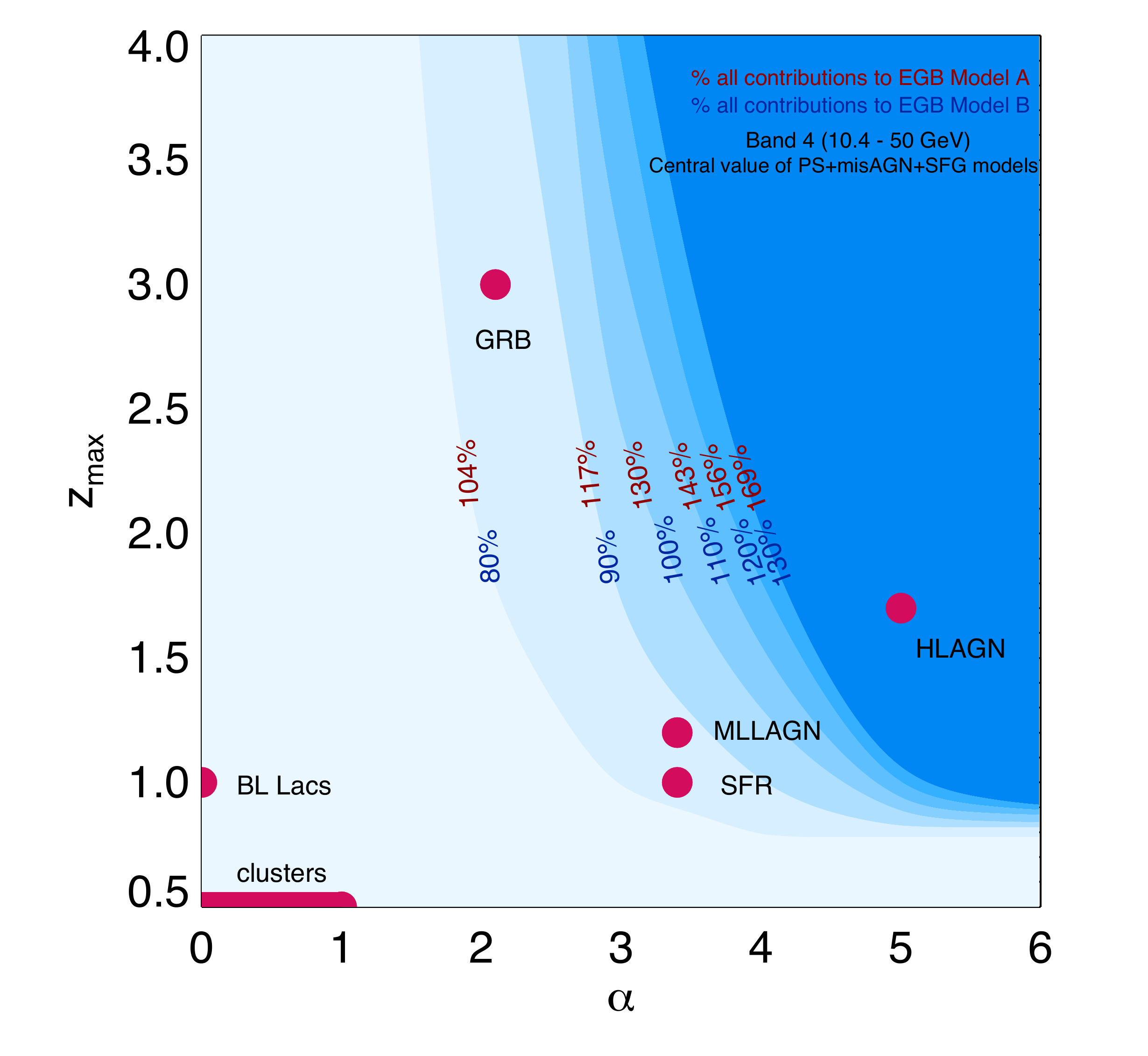}
\caption{Fermi-LAT constraints on EGCR source evolution in the case of our mixed-composition scenario and proton index $\beta=2.5$. 
The different colors show the percentage of the sum of all components (UHECR+PS+misAGN+SFG) to the EGB (Models A and B) in the 10.4--50~GeV energy band, in the ($\alpha$,~$z_{\max}$) parameter space, where $z_{\max}$ is the maximum redshift up to which sources experience a cosmological evolution in $(1+z)^{\alpha}$. Some possible EGCR sources \citep[see e.g.][for the references to the cosmological evolutions]{2016ApJ...822...56G} are shown. 
GRB: gamma-ray bursts. SFR: star-formation rate. MLLAGN: Medium-low-luminosity AGNs. MHLAGN: Medium-High-Luminosity AGNs. HLAGN: High Luminosity AGNs.
}
\label{fig:summary} 
\end{figure}

To determine whether a given EGCR source model  is  compatible with the $\gamma$-ray data, we need to take into account other known contributions to the EGB. Recently, A16 and Z16 showed that,  the EGB is dominated by (resolved and unresolved) PS, notably above $\sim$50 GeV, and estimated their contributions in six different energy bands, from 1~GeV to 2~TeV. These contributions are given in Table~\ref{table1} in terms of flux as well as percentage of the EGB, for both models A and B. 
While this PS flux is thought to be dominated by blazars, source populations with much smaller fluxes (thus mostly unresolved) may not be included in these estimates  \citep[see discussions in A16, Z16 and][]{Lisanti16}.
We thus consider in addition a possibly important contribution of star-forming galaxies (SFG) and misaligned active galactic nuclei (misAGN), based on the models by \citet{Inoue11} and \citet{Ack2012}.  Table~\ref{table1} gives their integrated fluxes and relative contributions to the EGB in the six energy bands considered by A16 and Z16. 
The SFG and misAGN $\gamma$-ray spectra are shown in Fig.~\ref{fig:EGB} (omitting the uncertainty bands for clarity).  Also shown 
 is the $\gamma$-ray spectrum arising from blazars, adapted from \citet{Ajello15}.
 This spectrum appears in good agreement with the PS contribution estimated by  A16 and Z16 over the whole energy range.

Turning now to include the contribution of the EGCRs to the $\gamma$-ray background we find that 
for  the GRB or non evolving scenarios, the sum of all components (UHECR, misAGN, SFG and blazars) never exceeds the total EGB, in the case of model B. 
In the case of model A, the sum is above the EGB. However, it falls below it if one adopts the $1\sigma$ lower bound on the misAGN+SFG+blazars contribution.

 \begin{figure}[t!]
 \centering
\includegraphics[width=\linewidth]{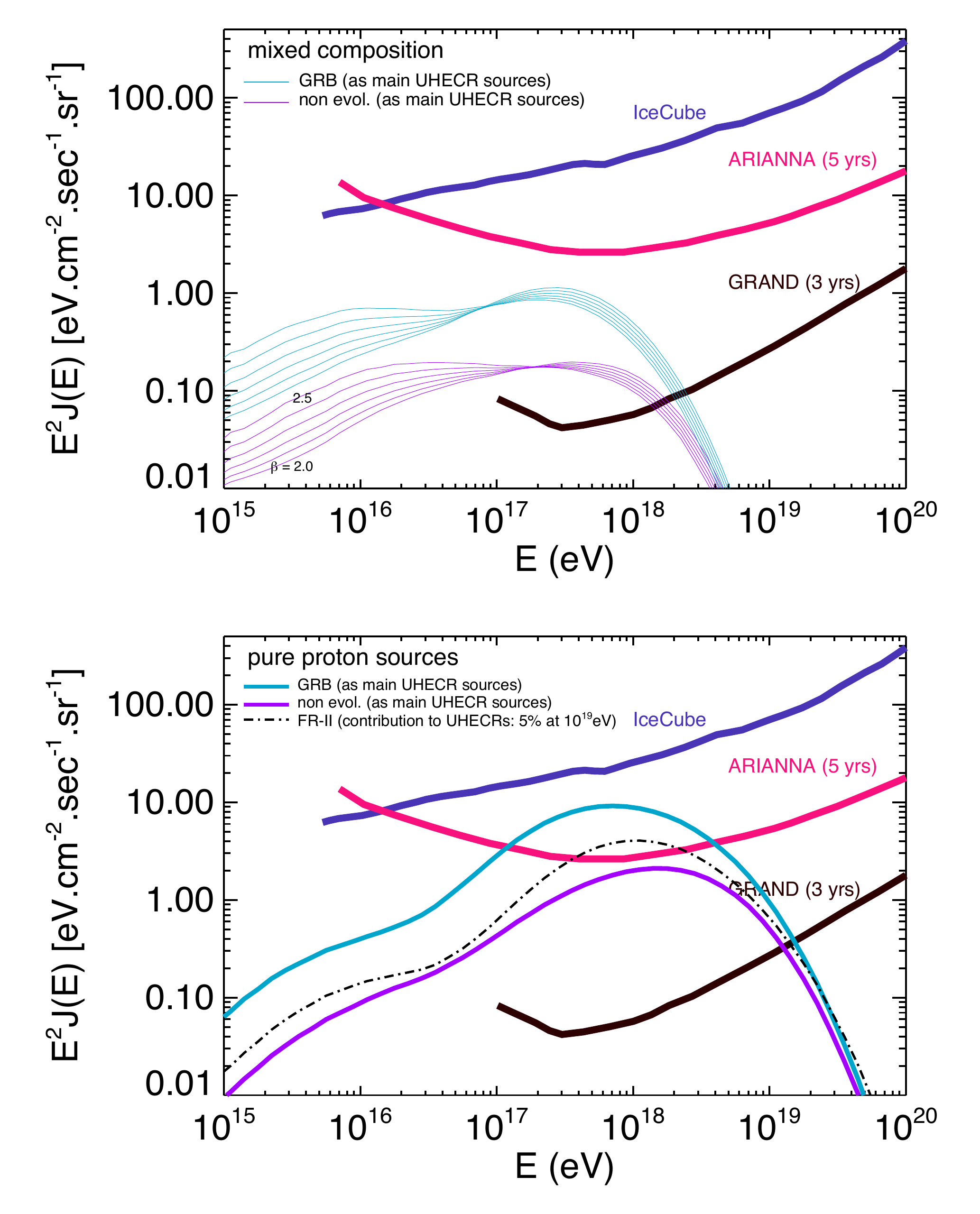}
\caption{ Upper panel: cosmogenic $\nu$ fluxes associated with our mixed-composition scenarios in the case of GRB-like evolution (blue) and non-evolving (violet) sources, compared with the current IceCube sensitivity \citep{Aartsen16} and the expected sensitivities of ARIANNA (5 years, 50 MHz option, \cite{Hallgren16}) and GRAND (3 years, \cite{Martineau15}). Lower panel: same, i) for 100\% proton scenarios compatible with the Fermi constraints (plain lines, same colour code), and ii) for a sub-dominant proton component (contributing 5\% of the UHECRs at 10~EeV) evolving as FR-II galaxies (dashed-dotted line) \citep{Wall05}.}
\label{fig:nu}
 \end{figure}

Table~\ref{table2} gives more details on the integrated $\gamma$-ray fluxes ($F_{\rm UHECR}$) contributed by the extragalactic UHECRs in the same six energy bands for which A16 and Z16 have estimated the PS contribution. We compare those contributions to the total EGB (Models A and B).
The 
percentages of these $\gamma$-ray fluxes from UHECRs 
to the EGB ($F_{\rm UHECR}/F_{\rm EGB}\times 100$), are shown for three different source evolutions (GRB, SFR, and non-evolving), and two different spectral index of the proton spectrum ($\beta=2.0$ and $2.5$).
We  also give the percentage of   
 the sum of PS, UHECR, misAGNs and SFGs  to the EGB.
 The sum of all components never exceeds the Fermi-LAT limits in the case of Model B, as already hinted by Fig.~\ref{fig:EGB}.
 
 The case of model A is less clear. The sum of the PS and $\gamma$-rays from UHECRs, without adding the more uncertain misAGNs and SFGs (part of which may already be included in the PS contribution estimated by A16 and Z16 anyway) respects  the observational constraints for all our models, as can be seen in the last three lines of Table~\ref{table2}. 
 However, should model A and the contribution of all the various PS (including misAGNs and SFGs) be confirmed, our calculations show a tension in the $\gamma$-ray and UHECR data, even in the case of the non-evolving scenario. 

Fig.~\ref{fig:summary} shows the allowed parameter space of different evolutionary scenarios. This estimate is based on the summed contribution of all components in the 10.4-50 GeV band, where the contribution from UHECRs is the largest. Only very strong evolutions are excluded by the current observations.\\\\

\section{Neutrino counterpart}

Fig.~\ref{fig:nu} shows the resulting $\nu$ spectra for different EGCR models, together with the sensitivity of current and planned experiments.
The mixed-composition models predict $\nu$ fluxes too low to be detected by IceCube \citep{Aartsen16} or ARIANNA \citep{Hallgren16}, even in the case of a GRB-like cosmological evolution.
They would require a sensitivity such as that expected for the GRAND observatory \citep{Martineau15} or CHANT satellite concept \citep{Neronov16}. For this reason, it is often considered that a possible future detection of cosmogenic $\nu$s by IceCube or ARIANNA would be a very strong argument against the mixed-composition UHECR models. 
Pure proton scenarios can indeed be seen on Fig.~\ref{fig:nu} to yield detectable fluxes, while still being allowed by the current IceCube limits and Fermi-LAT data. For these calculations, we assumed a pure proton $E^{-2}$ spectrum with an exponential cutoff at $E_{\rm max} = 60$~EeV (which is known to reproduce reasonably well the Auger spectrum above the ankle).

However,  it is interesting to note that when the $\nu$ flux is concerned there is a trade off  between the strength of the protonic UHECR sources  and its cosmic evolution.  Hence  a $\nu$ detection would {not necessarily} sign a pure proton scenario.  
An albeit  hypothetical at present, subdominant (less than $\sim$5--10\% of the UHECR flux) proton component with $E_{\rm max}\sim 10^{20}$~eV and a strong cosmological evolution, 
would  contribute a detectable  $\nu$ flux around $10^{18}$~eV (see Fig.~\ref{fig:nu}), 
while the bulk of the UHECRs would still be provided by sources with a mixed-composition and low proton $E_{\rm max}$.
Since this flux is much larger than that associated with the main mixed-composition EGCR component, a $\nu$ detection at that level may actually be the best way to reveal such a subdominant UHECR proton contribution. 

The contribution of such subdominant proton sources (correlated with their cosmological evolution) would also be constrained by their GeV-TeV $\gamma$-ray emission. This demonstrates the importance of multi-messengers studies, and their emerging power in constraining high-energy source models.

\section{Summary}

The UHECR model considered in G15b  
gives a coherent picture of the GCR-to-EGCR transition,
and  appears to be compatible with the Fermi-LAT measurements and the estimates of the PS contributions by A16 and Z16. 
It is compatible, with even more room for UHECRs, 
with the estimates of \citet{Lisanti16} for the PS contributions ($\sim$54\% and 68\% of the EGB Model A around 2 GeV and above 50 GeV, respectively).
Only very strong evolutions 
are excluded by the current observations.
The mixed-composition model appear to be less constrained by the Fermi-LAT 
 than the electron-positron dip (pure-proton) scenario \citep{Bere16, Supanitsky16, 2016ApJ...822...56G} that rules out SFR-like and stronger cosmological evolutions (see also \cite{Heinze16} for more radical conclusions on the dip model).

Our interpretation\footnote{We checked that the different interpretation did not originate from numerical discrepancies between the two studies.} differs from \citet{Liu16}. Considering only model A and 
a pure-proton composition at $10^{18}$ eV, these authors found a $\sim~1\sigma$ excess and therefore suggested the existence of a local overdensity of  $10^{18}$ eV proton sources. 
We find that these local proton sources are unnecessary.  Our UHECR model is consistent, within the current uncertainties of PS and Galactic foreground, with the EGB data. 


For the evolutionary models allowed by Fermi, the $\nu$s fluxes above $10^{17}$~eV associated with the mixed-composition scenario are well below the current  IceCube limits. These fluxes are 
within the reach only of the most sensitive planed $\nu$ observatories.
These fluxes could be outshined by the $\nu$s produced by hypothetical subdominant EGCR proton sources, with large enough $E_{\rm max}$ and cosmological evolution, thus making EeV $\nu$s a powerful probe for revealing the existence of trans-GZK proton accelerators, even if they do not dominate the observed UHECR flux.

Finally, we note that while the PS contributions are now understood to dominate the extragalactic $\gamma$-ray fluxes in the GeV-TeV range, the uncertainties on the different contributions \citep[notably for  sources other than blazars, see e.g][]{DiMauro13,Lacki14,Tamborra14} as well as on the Galactic foreground are still too large to efficiently constrain the cosmological evolution of UHECR sources. Since the $\gamma$-ray fluxes  associated  with mixed-composition UHECRs never exceed $\sim 20$\% of the EGB  (at least for source evolutions not significantly larger than SFR, see Table~\ref{table2}),  the EGB and its other contributions should be  { determined } to this level of precision in order to estimate whether a UHECR mixed-composition model is  excluded. 
Moreover, the Fermi-LAT estimates of the Galactic foreground are based on the GALPROP framework \citep{Strong00}. These calculations rely on several simplifying assumptions in particular in the description of the Galactic cosmic-ray source distribution or the magnetic halo, as well as on several {\it ad-hoc} parameters that are  tuned to reproduce cosmic-ray data. Alternative models 
\citep[e.g.][and references therein]{Nava17}
have been shown to fairly account for the primary-to-secondary ratios as well as some puzzling features in the observed $\gamma$-ray Galactic signal. 
These models have a smaller halo extension and would probably result in a lower Galactic foreground, leaving more room for EGCR contributions.

\begin{acknowledgments}
We thank Guillaume Decerprit for his important contribution to previous works, as well as Jean Ballet, Beno\^it Lott, Thierry Reposeur and Mario Bertaina for very useful discussions on the Fermi-LAT and KASCADE-Grande data. We thank David Eichler and Felix Aharonian for enlightening discussions. NG and TP acknowledges the I-CORE Program of the Planning and Budgeting Committee and The Israel Science Foundation (grant 1829/12), the advanced ERC grant TReX, and the Lady Davis foundation. 
\end{acknowledgments}

\end{document}